\documentclass[12pt,reqno]{article}
\usepackage{amsmath,hyperref,amssymb}
\usepackage{subfigure,graphicx,url}

\newcommand{\bea}{\begin{eqnarray}}
\newcommand{\eea}{\end{eqnarray}}
\newcommand{\be}{\begin{equation}}
\newcommand{\ee}{\end{equation}}
\newcommand{\eeq}{\end{equation}}
\newcommand{\beq}{\begin{equation}}

\allowdisplaybreaks \numberwithin{equation}{section}
\renewcommand{\Large}{\large} 

\DeclareSymbolFont{AMSa}{U}{msa}{m}{n}
\DeclareSymbolFont{AMSb}{U}{msb}{m}{n}
\DeclareMathSymbol{\fieldR}{\mathalpha}{AMSb}{"52}

\def\beq{\begin{equation}}
\def\eeq{\end{equation}}
\def\be{\begin{equation}}
\def\ee{\end{equation}}
\def\bea{\begin{eqnarray}}
\def\eea{\end{eqnarray}}

\begin{document}

\bigskip
\begin{center}
 {\Large\bfseries  Crystalline Scaling Geometries from Vortex Lattices}\\[3mm]

 \
 
 {\large Ning Bao and Sarah Harrison} \\[5mm]   
 
 {\small\slshape
  SITP, Department of Physics\\
 and\\
 Theory Group, SLAC \\
 Stanford University\\
 Stanford, CA 94305, USA \\
\medskip
 {\upshape\ttfamily ningbao@stanford.edu, sarharr@stanford.edu}
\\[3mm]}
\end{center}

\

\

\vspace{1mm}  \centerline{\bfseries Abstract}
\medskip

We study magnetic geometries with Lifshitz and/or hyperscaling violation exponents (both with a hard wall cutoff in the IR and a smooth black brane horizon) which have a complex scalar field which couples to the magnetic field. The complex scalar is unstable to the production of a vortex lattice in the IR. The lattice is a normalizable mode which is relevant (i.e. grows into the IR.) When one considers linearized backreaction of the lattice on the metric and gauge field, the metric forms a crystalline structure. We analyze the scaling of the free energy, thermodynamic entropy, and entanglement in the lattice phase and find that in the smeared limit, the leading order correction to thermodynamic properties due to the lattice has the scaling behavior of a theory with a hyperscaling violation exponent between 0 and 1, indicating a flow to an effectively lower-dimensional theory in the deep IR.

\bigskip
\newpage

\tableofcontents

\newpage


\section{Introduction}

A natural extension of the AdS/CFT correspondence is the holographic description of phases of strongly coupled QFTs with reduced symmetry. One such direction is the description of phases which break conformal invariance and exhibit dynamical scaling \cite{KLM,Taylor} and/or hyperscaling violation \cite{thetaglom,ogawa,HSS,edgar,theta}. A recent pursuit has been the more ambitious goal of describing phases with spatial anisotropy \cite{Bianchi,Nori,Vegh} or inhomogeneity\cite{ooguri,Maeda,gauntlettspace,Horowitz,Erdmenger,Cremonini}. Such systems are of interest on the field theory side because, e.g., impurities induce a variety of transport phenomena in condensed matter systems. These IR phases are dual to anisotropic and inhomogenous brane horizons; finding such novel brane solutions is interesting from the perspective of gravity alone.

This works builds on the recent results of \cite{vlat}, and the older studies in \cite{Maeda,Erdmenger,Johnson}. The motivation of \cite{vlat} was to study the phase structure of (2+1)-dimensional CFTs doped with a charge density. These types of systems can have nonperturbative instabilities associated with monopole operators which couple to the charge density. These operators are thought to lead to interesting solid phases in the IR \cite{Sachdevnew}.  Via electric-magnetic duality, the electric charge maps to a magnetic field, and the monopole operators map to electrically-charged operators coupled to the magnetic field. In \cite{vlat}, we work in this magnetic picture, and study the phase transition driven by the electrically-charged scalar. In the IR, we find the theory flows to a vortex lattice solution, which should have the properties of the solid phases suspected to appear in the IR of doped CFTs.

The holographic system we use is an $AdS_2\times R^2$ geometry (with a hard wall cutoff) supported by a magnetic field; this type of geometry is known to be the near-horizon region of extremal Reissner-Nordstrom black branes, as well as an endpoint of IR flows in more general Einstein-Maxwell-Dilaton (EMD) systems \cite{GKPT,HKW,Crem1,jtliu}. A complex scalar in the bulk coupled to the magnetic field obtains an instability at a critical value of the field with respect to the IR cutoff scale; beyond this critical value the scalar condenses into a vortex lattice. We also consider linearized backreaction on the metric and gauge field, where the small parameter is the distance away from the critical field. This results in a metric which has a crystalline form at leading order in the linearized approximation. In this work, we extend these results to bulk geometries with more general critical exponents, and geometries with a smooth horizon instead of a hard wall in the IR.

The outline of the rest of the paper is the following. In \S~\ref{sec:EMD}, we discuss the phase structure of the bulk EMD actions we will study, and review in detail the construction of the vortex lattice solution in \cite{vlat}. In \S~\ref{sec:gen}, we extend these vortex lattice solutions to metrics with general values of the critical exponents $z$ and $\theta$. We analyze the effects of the vortex impurities on thermodynamic quantities in \S~\ref{sec:thermo}, and find that the dimension $\Delta$ of the scalar operator determines the scaling of most thermodynamic quantities. In addition, we compute entanglement entropy of the lattice in the smeared limit and find that the lattice correction scales as a power law, $\ell^{2\Delta}$, where $\ell$ is the width of the entanglement strip. We find evidence from scaling of free energy of the field theory and scalar curvature in the bulk that the vortex impurities act as (0+1)-dimensional theories with an effective hyperscaling violation exponent, where $d-1<\theta<d$ in the $AdS_2$ case, an example of the novel phases found in \cite{theta}. Finally, in \S~\ref{sec:discussion}, we end with a discussion of our results, their relation to phases of doped CFTs, and open questions. 

\section{Phases of EMD systems}\label{sec:EMD}
In this section, we discuss the phase structure of EMD actions in the bulk with action of the form, 
\beq\label{eq:zthet_action}
S=\int d^4 x \sqrt{-g} \left(R -\frac{1}{2}\left|\nabla \phi \right|^2-\frac{1}{4}F^2e^{2\alpha\phi}-2\Lambda e^{2\delta\phi}\right),
\eeq
where $\Lambda < 0$, and the gauge coupling is controlled by the dilaton field, $g_\phi \sim e^{-\alpha\phi}$. These systems can be supported by an electrically---or, via S-duality, magnetically---charged extremal black brane with near-horizon metric,
\be\label{eq:metric}
ds^2=L^2r^\theta\left(-\frac{dt^2}{r^{2z}}+\frac{dr^2}{r^2}+\frac{dx^2+dy^2}{r^2}\right),
\ee
where $z$ is the dynamical critical exponent, $\theta$ is the hyperscaling violation exponent, and $r\to 0$ is the UV boundary of bulk spacetime. In the IR region where the bulk is described by equation (\ref{eq:metric}), the stress-energy in the dual field theory scales with the symmetries of this metric. 

For electrically charged branes, the gauge field has the form $F\sim e^{-2\alpha\phi} r^\gamma Q_e dt\wedge dr$, with $\gamma$ set by the choice of $\alpha, \delta$ in the action, whereas for magnetic branes the gauge field has the form $F\sim Q_m dx\wedge dy$. In the former, $Q_e$ is electric charge, while $Q_m$ is magnetic charge. In both cases, the dilaton field takes the form $\phi\sim \phi_0 +k\log r$ with $k$ a constant.

As discussed in \cite{GKPT,HKW}, the log-running form of the dilaton in the above equations causes singular behavior as one flows to the IR, $r\to \infty$. In the electric geometries, the sign of $k$ is positive, and the theory flows toward weak coupling in the IR, i.e. $g_\phi\to 0$; $\alpha'$ corrections become important in the IR, and one must take into account higher curvature terms in the action in order to determine the full RG flow \cite{jtliu}. On the other hand, in the magnetic system, the sign of $k$ is negative, and the theory becomes strongly coupled in the IR, indicating quantum corrections are becoming important.

It has recently been shown that after including quantum corrections in systems with Lifshitz scaling \cite{GKPT}, the dilaton is stabilized in the deep IR and the geometry flows to $AdS_2 \times R^2$ \cite{HKW}. This story was generalized for metrics with a hyperscaling violation exponent in \cite{Crem1}, and for electric systems with higher derivative terms in \cite{jtliu}.

Though there is no charged solution in pure $AdS_4$, it is possible to achieve $AdS_4$ asymptotics in the UV \cite{GKPT}; in these cases the dilaton is stabilized at a constant value in the UV, $\phi_{UV}$, and the $F^2$ term becomes vanishingly small compared to the cosmological constant, allowing for restoration of the full conformal invariance at $r=0$. 

Now, let's discuss the scaling solution of equation (\ref{eq:metric}) a bit further. The scalar curvature in these geometries behaves as $R\sim r^{\theta}$; in particular, for $\theta > 0$ it diverges in the far IR, as $r\to \infty$. However, in the hard wall case (like in \cite{vlat}), we can cut off the geometry at some $r=r_0<\infty$, so we can pick an $r_0$ such that the curvature remains small.\footnote{In addition, there is the above-mentioned fact that the dilaton is stabilized in the IR and there is a re-emergent $AdS_2\times R^2$ geometry \cite{HKW,Crem1,jtliu}.}  

The NEC give constraints on the physically allowed values of $z$ and $\theta$ \cite{theta}. For theories with $\theta=0$, $z\geq 1$. For Lorentz-invariant theories, $z=1$, and either $\theta\leq 0$ or $\theta \geq 2$ (though the latter are suspected to be unstable.) Finally, for $\theta=d-1=1$, which is conjectured to be dual to field theories with a Fermi surface \cite{ogawa,HSS}, $z\geq {3\over 2}$.

We will primarily be interested in the system with the magnetically charged brane. Via electric-magnetic duality, this is dual to a CFT doped with a finite charge density in the electric picture. We can write the critical exponents and $\Lambda, k, L$ in terms of the parameters $\alpha$ and $\delta$ as,
\bea
k=\frac{\theta-4}{2\alpha}=-\frac{\theta}{2\delta}\\
\Lambda=-\frac{(1+2 \alpha  (\alpha -\delta )) (1+2 (\alpha -\delta ) \delta ) (1+(\alpha -\delta ) (3 \alpha +\delta ))^2}{\left(\alpha ^2-\delta ^2\right)^4}\\
z=\frac{1+\alpha ^2+2 \alpha  \delta -3 \delta ^2}{\alpha ^2-\delta ^2}\\
\theta=\frac{4 \delta }{\alpha +\delta }\\
L=\frac{\alpha ^2-\delta ^2}{\sqrt{\left(2+4 \alpha  \delta -4 \delta ^2\right) (1+(\alpha -\delta ) (3 \alpha +\delta ))}},
\eea
where we have set $\phi_0=0$.

 In other cases, it is sometimes convenient to write the above parameters in the following form:
\bea
\alpha=\frac{\theta-4}{2\sqrt{-4-2z(-2+\theta)+\theta^2}}\\
\delta=\frac{\theta}{2\sqrt{-4-2z(-2+\theta)+\theta^2}}\\
\Lambda=-(z-1)(1+z-\theta)(2+z-\theta)^2\\
k=\sqrt{-4-2z(-2+\theta)+\theta^2}\\
L=\frac{1}{\sqrt{2(z-1)(2+z-\theta)}}.
\eea
We stick to $\theta\geq0$, as it turns out there will not be a stable lattice solution if $\theta <0$, (even though this is in general allowed by the NEC.) Note that we have set $Q=1$; we will choose this value of $Q$ in all future computations as well.

\subsection{Inclusion of ``monopole" operator}
Now consider including a monopole operator in these systems. Monopoles in the electric picture are dual to electrically charged scalars in the magnetic picture. We include the effects of this field by perturbing the action with a term of the form,
\be
\delta S=\int d^4x\left ( -|\nabla_\mu \psi|^2-m^2|\psi^2|-\lambda |\psi|^4\right ),
\ee
where $\nabla_\mu=\partial_\mu+iA_\mu$. This action combined with that of equation (\ref{eq:zthet_action}) must satisfy the Einstein equations,
\be
R_{\mu\nu}-\frac{(R-2\Lambda e^{2\delta\phi})}{2}g_{\mu\nu}=T_{\mu\nu},
\ee
where the total stress-energy tensor is
\bea\nonumber
T_{\mu\nu}&=&-{g_{\mu\nu}\over 2}\mathcal{L}_{mat}+{1\over 2}e^{2\alpha\phi}F_{\mu\lambda}F_\nu^\lambda+e^2A_\mu A_\nu|\psi|^2+
{1\over2}(\partial_\mu\psi\partial_\nu\psi^*+ie\psi(A_\mu\partial_\nu+A_\nu\partial_\mu)\psi^*\\
&&+h.c.)+\frac{1}{2}\partial_\mu\phi\partial_\nu\phi
\eea
with Lagrangian
\be
\mathcal{L}_{mat}={1\over 4}e^{2\alpha\phi}F^2+|\nabla_\mu \psi|^2+\frac{1}{2}(\partial_\mu \phi)^2+m^2|\psi|^2+\lambda |\psi|^4,
\ee
and equations of motion for $\phi$
\be
\partial_\mu(\sqrt{-g}\partial^\mu\phi)=\sqrt{-g}\left(4\delta \Lambda e^{2\delta\phi}+{\alpha\over 2} e^{2\alpha\phi}F^2\right),
\ee
$\psi$,
\be
\partial_\mu(\sqrt{-g}\nabla^\mu\psi)=-\sqrt{-g}(iA^\mu\nabla_\mu\psi-m^2\psi-2\lambda|\psi|^2\psi),
\ee
and $A_\mu$,
\be
{1\over\sqrt{-g}}\partial_\mu(\sqrt{-g} e^{2\alpha\phi}F^{\mu\nu})=i(\psi\partial^\nu\psi^*-\psi^*\partial^\nu\psi)+2A^\nu|\psi|^2.
\ee
In the UV we expect the lowest energy configuration to be $\psi=0$; however, by tuning the magnitude of the magnetic field, one can produce an instability which causes it to condense to a vortex lattice state. 
In \cite{vlat}, this phase transition was explored directly in the IR $AdS_2\times R^2$ background described in the previous section. 

We will expand perturbatively in a small parameter $\epsilon$ around the solution $\psi=0$ with background gauge field of the form
\begin{equation}
\label{gaugechoice}
A_x = Q_c  y, A_y = 0.
\end{equation}
The scalar field in the competing vortex phase  will itself be of order $\epsilon$.
For fixed $r_0$ and boundary conditions (to be discussed below), we will choose $Q_c$ to be just at the onset for the transition to vortex formation. At this critical value of the magnetic field, the $\psi=0$ solution will be degenerate with a vortex lattice solution. As we increase the magnetic field to slightly above its critical value, $\psi=0$ will no longer be the preferred solution, and the vortex lattice will be preferred.  As is familiar, the onset of the transition is signalled by the emergence of a purely normalizable solution for 
$\psi$ that respects the IR boundary conditions.


We can parametrize the backreaction of the scalar on the gauge sector through a perturbative expansion in the distance away from the critical field.
The scalar will have the form
\be\label{eq:psi_expansion}
\psi(r,x,y)=\epsilon\psi_1(r,x,y)+\epsilon^3\psi_3(r,x,y)+\ldots
\ee
and the gauge field will have the form
\bea\label{eq:A_expansion}
A_x(r,x,y)&=&Qy+\epsilon^2a^x_2(r,x,y)+\ldots,\\
A_y(r,x,y)&=&\epsilon^2a^y_2(r,x,y)+\ldots,
\eea
with $A_t=A_r=0$. When we consider lattice solutions which are periodic in $x$ and $y$, the backreaction of $\psi$ on the gauge field will require both $A_x$ and $A_y$ to be nonzero at $O(\epsilon^2)$, with both $x$ and $y$ dependence. 

A similar statement holds for the metric at $O(\epsilon^2)$.
Our metric ansatz, to $O(\epsilon^2)$, will be
\be
ds^2=L^2\left \{{1\over r^2}((-1+\epsilon^2 a(r,x,y))dt^2+(1+\epsilon^2 a(r,x,y))dr^2)+(1+\epsilon^2 b(r,x,y))(dx^2+dy^2)\right \}.
\ee
Because at zeroth order in epsilon the $AdS_2\times R^2$ metric is exactly supported by the magnetic field (i.e. the gauge field is not a probe), we find it necessary to include metric backreaction once we backreact on the gauge field.  This distinguishes our situation from that considered in e.g. \cite{Erdmenger}.

The radial magnetic field will be 
\be
B_r=Q+\epsilon^2(\partial_ya^x_2(r,x,y)-\partial_xa^y_2(r,x,y))
\ee

In general, when we backreact on the magnetic field, we may expect there to be a non-normalizable piece at order $\epsilon^2$, i.e. $A_x(r\to 0)= (Q+\bar Q\epsilon^2)y$. This shifts the naive critical value of the field at the transition. However, because the critical point is actually only dependent on a dimensionless combination of $Q$ and $r_0$, we can (and will) impose that there is no non-normalizable correction to the gauge field in our backreacted solutions. That is, we will set $\bar Q=0$. The value of the critical point will still have an $O(\epsilon^2)$ shift; it will manifest itself as an $O(\epsilon^2)$ shift in the location of the hard wall, $r_0\to r_0+\bar r_0\epsilon^2$. These two scenarios are equivalent; in both cases we should think of the backreaction of $\psi$ on the metric and gauge field as inducing a shift in the dimensionless parameter which controls the critical point at $O(\epsilon^2)$.

\subsection{Engineering a lattice transition}
Here we describe the form of the vortex solution found in \cite{vlat}.The only nontrivial equation of motion at $O(\epsilon)$ is the $\psi$ equation of motion. This equation has a separable solution, $\psi(r,x,y)=\rho_0\rho(r)g(x,y)$, with $\rho_0$ a constant, independent of the form of the metric. Since we are doing linearized backreaction, this is always the case; this will prove useful when we generalize to metric backgrounds which are more complicated than $AdS_2\times R^2$. 

For an $AdS_2\times R^2$ background, $\rho(r)$ takes a simple power law form, $\rho(r)=r^{\alpha}$, where $\alpha=\frac{1}{2}\left(1+\sqrt{1+4Q+4m^2L^2}\right)$ where ${1\over 2}<\alpha<1$ and $m^2$ can be negative (but above the BF bound). This means $\psi$ is a relevant operator, normalizable at the boundary, and the strength of the lattice grows into the IR. The shape of the vortex lattice appears in the function $g(x,y)$, which can be written in terms of a Jacobi-theta function as 
\be
g(x,y)=e^{-Qy^2\over 2}\theta_3\left ({\sqrt{Q}(x+iy)\over v_1},{2\pi i\over v_1^2}\right ),
\ee
where the theta function is defined as 
\be
\theta_3(v,\tau)\equiv \sum_{l=-\infty}^\infty q^{l^2}z^{2l},~q\equiv e^{i\pi\tau},~z\equiv e^{i\pi v},
\ee
and obeys the relations $\theta_3(v+1,\tau)=\theta_3(v,\tau)$ and $\theta_3(v+\tau,\tau)=e^{-2\pi i(v+\tau/2)}\theta_3(v,\tau)$.
 (This form of $g(x,y)$ is for a rectangular lattice with sides of length $v_1$ and ${2\pi \over v_1}$. Each unit cell of the lattice has area ${2\pi \over Q}$, containing exactly one flux quantum.For more details, see \cite{vlat}.)

At second order we must solve the gauge field equations of motion and the Einstein equations. The manifest double periodicity in the $x,y$ plane of the lattice solution allows us to expand the functions $a_2^x,a_2^y, a,$ and $b$ as a function of $r$ times a double Fourier series in $x,y$:
\be\label{eq:fourier}
f_i(r,x,y)=\rho_0^2r^{2\alpha}\sum_{k,l} e^{{2\pi i k\sqrt{Q}x\over v_1}}e^{ilv_1\sqrt{Q}y}\tilde f_i(k,l),
\ee
where $f_i=a_2^x,a_2^y,a,b$. This renders all the equations of motion at $O(\epsilon^2)$ algebraic. 

We cut off the geometry in the IR with a hard wall at a radial value $r=r_0$; therefore, we must specify a consistent set of boundary conditions at the wall. We can think of this radial slice as gluing together two copies of the geometry in the IR, such that all of the fields have the same form as a function of $|r-r_0|$ \cite{RS}. In order to satisfy the Einstein equations at $r=r_0$, there must be a source of stress-energy localized to the wall. Specifying the form of this stress energy will yield a unique solution for $\alpha$, the scaling dimension of the lattice, as a function of $r_0$, the location of the wall. We assume that there is some mechanism, possibly similar to \cite{GW}, stabilizing the hard wall such that we can exclude the possibility of fluctuations of the wall.

Unlike the mechanism of the original constructions of holographic superconductors in a black hole background \cite{gubser,hhh}, (a scalar mass which is stable in the UV goes below the BF bound \cite{BFB} in the IR and causes the scalar to condense,) we have a fixed background geometry, and we tune the stress-energy at the wall and the location of the wall in order to force a scalar mode to become unstable. However, the principle is the same; in both cases a scalar mode becomes unstable at a critical energy scale and obtains a vev. Of course one could thicken the IR wall to obtain other IR geometries, e.g. with smooth horizons, where the same physics of lattice formation occurs.

\section{Generalization to Lifshitz and hyperscaling violation}\label{sec:gen}
In this section we consider the transition to a vortex lattice in a magnetic background with nontrivial Lifshitz and/or hyperscaling violation exponents. This occurs when $\psi$ becomes unstable above the scale at which the solution of equation (\ref{eq:zthet_action}) crosses over to $AdS_2\times R^2$. This amounts to finding a vortex lattice solution in the background metric (\ref{eq:metric}). We will follow the same method as in \cite{vlat}, performing a linearized expansion around the critical magnetic field, so we will only highlight the differences from the previous solution. In this section, we do this in the case of a hard wall, but in the next section when we introduce a horizon, we will see an example of a more physical mechanism to produce an instability.

\subsection{Scaling of lattice condensate}\label{sec:psi_sol}
As mentioned before, at $O(\epsilon)$, the $\psi$ equation of motion is separable, independent of the metric, so we may choose the same $g(x,y)$ lattice solution as in the $AdS_2\times R^2$ case. The only thing that changes is the radial dependence of $\psi$. We find that $\rho(r)$ must satisfy the equation of motion
\be\label{eq:rho_eom}
\rho''(r)+\frac{\theta-z-1}{r}\rho'(r)-(m^2L^2r^{\theta-2}-Q)\rho(r)=0,
\ee
for general $z$ and $\theta$.

We can solve for $\rho (r)$ analytically for $\theta=0$ and $\theta=1$:
\beq
\theta=0: \rho(r)=r^{\frac{2+z}{2}}K_n(\sqrt{Q}r), 
n=\frac{1}{2} \sqrt{4L^2m^2+(z+2)^2}
\eeq
\beq
\theta=1:\rho(r)=e^{-\sqrt{Q} r} r^{1+z}U\left(\frac{L^2 m^2}{2 \sqrt{Q}}+1+{z\over 2},2+z,2 \sqrt{Q} r\right)
\eeq
where $K_n$ is the modified Bessel function of the second kind, and $U$ is the confluent hypergeometric function of the second kind. We have chosen solutions which are normalizable at the boundary, i.e. $\rho(r)=0$ as $r\to 0$. Setting $Q=1$ and using the solutions for the other parameters from \S~\ref{sec:EMD}, we get
\beq
\theta=0: \rho(r)=r^{\frac{2+z}{2}}K_n(r), 
n=\frac{1}{2} \sqrt{{2m^2\over (z+2)(z-1)}+(z+2)^2}
\eeq
\beq
\theta=1:\rho(r)=e^{- r} r^{1+z}U\left(\frac{m^2}{4(z^2-1)}+1+{z\over 2},2+z,2 r\right),
\eeq
if $z\neq 1$.

Note that the parameters in the solution for $\rho(r)$ give a lower bound on the mass of the scalar, which may be negative. We identify this as the BF bound \cite{BFB} for theories with a complex scalar in a gravitational background with general $z$ and/or $\theta$.  For $\theta =0$, the bound is
\be
m^2\geq -\frac{(z+2)^3(z-1)}{2},
\ee
and for $\theta =1$,
\be
m^2\geq -2(z+2)(z^2-1).
\ee
For values of $m$ below these limits, there is no physical solution for $\psi$.

For other values of $\theta$, (i.e. $\theta \neq 0,1$,) one must resort to numerics or matching asymptotic expansions. We will not delve into that here because we can still compute the radial scaling of $\psi$ near the boundary without knowing the full analytic form of $\psi(r)$ for all $r$. Moreover, this dimension is what primarily determines the physics of the vortex lattice which we discuss in the following sections. We find that, near the boundary, the radial part of $\psi$ scales as $\psi\sim r^\Delta$ with,\footnote{As mentioned before, this analysis only holds for values $\theta \geq 0$.}
\be\label{eq:dim}
\Delta_{\theta\neq 0} =2+z-\theta~~ \text{and}~~ \Delta_{\theta=0} = 1+{z\over 2}+\frac{1}{2} \sqrt{{2m^2\over (z+2)(z-1)}+(z+2)^2}.
\ee
Note that the values of $\Delta$ match when we take the limit $m\to 0$ in $\Delta_{\theta=0}$ and set $\theta=0$ in  $\Delta_{\theta\neq 0}$. Once we have a value $\theta>0$ , the $m^2$ term becomes subdominant in equation (\ref{eq:rho_eom}), and so the mass does not enter into the leading power of $\rho(r)$ in this case. This is the same as what was found in \cite{theta}.
The dimension in equation (\ref{eq:dim}) will govern the leading order behavior of the free energy, thermal entropy, and entanglement entropy in the lattice phase.\footnote{Of course, we would need the full solution $\psi_1(r,x,y)$ to calculate correct coefficients in these quantities.} We will come back to this in \S~\ref{sec:thermo}.

\subsection{Transition to the lattice phase}

Finally, we must address the boundary conditions by finding a potential localized at the wall which supports our solution. Placing a potential at the wall allows us to induce a transition, as there will be a critical value of $r_0$ (the location of the wall) at which there first appears a normalizable mode for $\psi$ which satisfies the boundary conditions required by the potential. 

Let's consider the action at the wall which enters at zeroth and first order in $\epsilon$--these set the IR boundary conditions for the dilaton and $\psi$. We will add the term,
\be
S_{wall}=\int\limits_{r=r_0} d^3x\sqrt{-h}\left({1\over 2}\delta m_\phi^2\phi^2 +\delta m_\psi^2|\psi|^2\right),
\ee
to the action, where $h_{\mu\nu}$ is the induced metric at $r=r_0$, and $\delta m_\phi, \delta m_\psi$ are localized shifts in the masses of the fields. The nonzero contributions to the equation of motion when integrated over the wall are
\be
-\int_{r_0-\epsilon}^{r_0+\epsilon} dr \sqrt{-g}g^{rr}\phi''=\int_{r_0-\epsilon}^{r_0+\epsilon}  dr \sqrt{-h}~\delta m_\phi^2\phi\delta(r-r_0)
\ee
and
\be
-\int_{r_0-\epsilon}^{r_0+\epsilon} dr \sqrt{-g}g^{rr}\psi''=\int_{r_0-\epsilon}^{r_0+\epsilon}  dr \sqrt{-h}~\delta m_\psi^2\psi\delta(r-r_0),
\ee
the first of which gives the result 
\be
\delta m_\phi^2 =\frac{2L}{(\theta-z-2)r_0^{\theta\over 2}\log r_0},
\ee
and the second of which gives a similar but messier result for $m_\psi^2$. Note that as $\theta$ will always be less than $z+2$, the mass-squared value at the wall is negative. The fact that the stress-energy at the wall is negative is consistent with what was found in \cite{vlat}. The quantity $r_0$ is measured in units of a crossover scale, $r_F$, below which the zeroth order hyperscaling solution is valid. One goes deeper into the IR by increasing $r$; therefore, in the above equation we have $r_0 >1$.

We will need to add additional terms to $S_{wall}$ when we consider the equations of motion at $O(\epsilon^2)$ in the next section.

\subsection{Linearized backreaction on the metric}\label{sec:second_order}
The vev for $\psi$ sources corrections to the gauge field, metric, and dilaton at $O(\epsilon^2)$. A general Lifshitz exponent breaks the $AdS_2$ symmetry between $t$ and $r$; therefore, the $O(\epsilon^2)$ corrections to the metric will be different for $g_{tt}$ and $g_{rr}$.  This gives us three  metric functions for the case $z\neq 0$.
Therefore, we correct the metric at this point to,
\be
ds^2=L^2r^\theta\left(-r^{-2z}\left(1-\epsilon^2 a(r,x,y)\right)dt^2+\frac{dr^2}{r^2}\left(1+\epsilon^2 a_1(r,x,y)\right)+\frac{(dx^2+dy^2)}{r^2}\left(1+\epsilon^2 b(r,x,y)\right)\right).
\ee

We also must expand the dilaton field,
\be
\phi(r)=\phi_0(r)+\epsilon^2\phi_2(r,x,y)+\ldots
\ee
where $\phi_0(r)=k\log r$ with $k$ given as in \S~\ref{sec:EMD}.

At $O(\epsilon^2)$ we now have the Einstein equations, gauge field equations of motion, and dilaton equation of motion to solve, each of which is sourced by the $O(\epsilon)$ solution for $\psi$. There are six unknown functions, $a(r,x,y), a_1(r,x,y), b(r,x,y), a_2^x(r,x,y), a_2^y(r,x,y),$ and $\phi_2(r,x,y)$. All of these functions will scale as $\sim r^{2\Delta}$ with $\Delta$ given by equation (\ref{eq:dim}) near the boundary, because they are sourced by terms $\sim |\psi|^2$. However, we'll need to turn to the equations of motion to find exact solutions for these functions.

Here we'll outline a detailed method for obtaining (in some cases, analytic) solutions with consistent boundary conditions at $r=r_0$ for these functions.
\begin{enumerate}
\item Expand each unknown function in a double Fourier series with the same periodicities as that of $\psi(x,y)$\footnote{From this point on, we set $Q=1$ for simplicity.},
\be\label{eq:exps}
f_i(r,x,y)=\rho_0^2\sum_{k,l}e^{{2\pi ikx\over v_1}}e^{ilv_1y}e^{-{k^2\pi^2\over v_1^2}-i\pi kl-{l^2v_1^2\over 4}}\tilde f_i(r,k,l),
\ee
where $f_i\in\{a,a_1,b,a_2^x,a_2^y,\phi_2 \}$. Notice that unlike in the $AdS_2\times R^2$ case, the $r$-dependence does not decouple from the $(x,y)$-dependence, so each Fourier coefficient is a function of $r$. However, the coefficients are exponentially suppressed as a function of $k^2$ and $l^2$, so we only need to include a few terms to have a very accurate numerical approximation for the $f_i(r,x,y)$. 

We plug these forms, as well as the solution for $\psi$ found in \S~\ref{sec:psi_sol}, into the equations of motion to yield six unique coupled ODEs for the six functions of $r$ for each $(k,l)$ value. The detailed form of these differential equations is given in the appendix.

\item Decouple the equations to get solutions for the $\tilde f_i(r,k,l)$. We show this process in detail for the zero modes, $\tilde f_i (r,0,0)$ below. As was found in \cite{vlat}, the zero modes for $a_2^x$ and $a_2^y$ vanish. (See the appendix for the detailed equations of motion.) We have left four equations (three Einstein and one dilaton) and the four unknown zero-mode functions  $\tilde a(r,0,0),\tilde a_1(r,0,0),\tilde b(r,0,0)$ and $\tilde \phi_2(r,0,0)$. 

Just for the rest of this section, we shorten the notation for the $r$-dependent zero-mode coefficients as $\tilde a(r,0,0)\equiv a(r)$, $\tilde a_1(r,0,0)\equiv a_1(r)$, $\tilde b(r,0,0)\equiv b(r)$, and $\tilde \phi_2(r,0,0)\equiv \phi(r)$.
It is convenient to introduce the auxiliary variables, keeping in mind the values of $k,Q,\Lambda$ and $L$ from above:
\bea
f(r)=b(r)+\frac{\sqrt{(\theta-2)(2-2z+\theta)}}{2-\theta}\phi(r)\\
g(r)=a_1(r)-\frac{\theta \phi(r)-2r\phi'(r)}{\sqrt{(\theta-2)(2-2z+\theta)}}\\
h(r)=-\frac{a(r)}{2+z-\theta}+\frac{(2z-\theta)\phi(r)}{2(2+z-\theta)\sqrt{(\theta-2)(2-2z+\theta)}}
\eea
These yield the more tractable set of coupled differential equations
\bea\nonumber
\frac{8(1+z-\theta)}{-2+2z-\theta}f(r)+\frac{2(1+z-\theta)}{(z-1)(2+z-\theta)}r f'(r)-\frac{2}{(z-1)(2+z-\theta)}r^2 f''(r)\\\nonumber=\frac{v_1\left(m^2r^{\theta}+2r^2(z-1)(2+z-\theta)\right)\rho(r)^2}{2\sqrt{\pi}(z-1)^2(2+z-\theta)^2},\\
\nonumber
g(r)-\frac{rg(r)}{2+z-\theta}-\frac{2(z-1)(6+2z-3\theta)}{(\theta-2)(2-2z+\theta)}f(r)+\frac{2(z-1)}{(\theta-2)(2+z-\theta)}f'(r)\\\nonumber=\frac{v_1\left(m^2r^{\theta}+2r^2(z-1)(2+z-\theta)\right)\rho(r)^2}{4\sqrt{\pi}(z-1)(2+z-\theta)^2(\theta-2)}-\frac{v_1r^2\rho'(r)^2}{2\sqrt{\pi}(2+z-\theta)(\theta-2)},\\\nonumber
\frac{2(z-1)(\theta-4)}{(2z-2-\theta)(\theta-2)}f(r)+\frac{rf'(r)}{2+z-\theta}+g(r)-\frac{rg'(r)}{4+2z-2\theta}+rh'(r)=0,\\
\nonumber
\frac{r(2z-\theta)}{(2+z-\theta)\sqrt{(2-2z+\theta)(\theta-2)}}\phi'(r)=-\frac{2(z-1)(4z-3\theta)}{(-2+2z-\theta)(2z-\theta)}f(r)+g(r)+\\\nonumber
\frac{r^2v_1\left(\rho(r)^2-\rho(r)'^2\right)}{2\sqrt{\pi}(2+z-\theta)(2z-\theta)}+\frac{r(\theta-z-1)g'(r)}{(2+z-\theta)(2z-\theta)}+4rh'(r)-2r^2h''(r),
\eea
which can be solved recursively by putting in the solution to $\rho(r)$ into the first equation and proceeding by putting the solution for the previous equation into the equation below; i.e. first we solve for $f(r)$, and then plug this solution into the second equation to get $g(r)$, $\ldots$. We choose the mode that is normalizable at the boundary. The solutions to these are analytic, however extremely long. The substitutions and equations simplify a bit for specific values of $z$ and $\theta$, say $z=2$ and $\theta=0$:
\bea
f(r)=b(r)+\phi(r)\\
g(r)=a_1(r)-r\phi'(r)\\
h(r)=\frac{2\phi(r)-a(r)}{8}.
\eea
For these values, the above equations reduce to the set
\bea\nonumber
12f(r)+\frac{3}{2}rf'(r)-\frac{1}{2}r^2f''(r)=\frac{\rho(r)^2v_1\left(8r^2+m^2\right)}{32\sqrt{\pi}},~\\\nonumber
g(r)-\frac{1}{4}rg'(r)-5f(r)-\frac{1}{4}rf'(r)=-\frac{r^2v_1\left(\rho^2(r)-\rho'^2(r)\right)}{16\sqrt{\pi}}-\frac{m^2v_1\rho^2(r)}{128\sqrt{\pi}~},\\\nonumber
2f(r)+\frac{1}{4}rf'(r)-g(r)-\frac{1}{8}rg'(r)+rh'(r)=0,~\\\nonumber
r\phi'(r)=-4f(r)+2g(r)+\frac{r^2v_1\left(\rho^2(r)-\rho'^2(r)\right)}{16\sqrt{\pi}}-\frac{3}{8}rg'(r)+8rh'(r)-r^2h''(r).~
\eea
Again, these are analytically solvable, though with long solutions not worth reproducing here. 

Finally with the solution for $\phi(r)$ at hand, we get solutions for $a,a_1,$ and $b$ by plugging back in to the above equations.

Then we must repeat this for the $r$-dependent functions which are coefficients of the modes of higher $(k,l)$ in the Fourier expansion. One might worry that these coupled ODEs are not generically solvable for all values of $(k,l)$, and thus there is not a good backreacted solution in the linearized approximation. Fortunately, this is not the case. As we show explicitly in the appendix, at a particular value of $(k,l)$, the equations of motion reduce to six linearly independent coupled second order ODEs of six functions, sourced by terms $\sim |\psi_1|^2$. This problem is generically solvable. 

\item The last thing we must do to find a consistent solution is fix the IR boundary conditions. We do this just as in \cite{vlat}, by allowing for a source of stress-energy at $r=r_0$.

At $O(\epsilon^2)$ we need to consider the gauge field equations of motion, the dilaton equation, and the Einstein equations integrated across the wall. We will now consider the following action at the wall,
\be
S_{wall}=\int\limits_{r=r_0} d^3x\sqrt{-h}\left \{{1\over 2}\delta m_\phi^2\phi^2+\delta m_\psi^2|\psi|^2+ \epsilon^2A_\mu J_w^\mu + \epsilon^2(T_w)_\mu^{~\mu}\right \},
\ee
where we have added a current $J_w^\mu$ which couples to the gauge field, as well as a source of stress-energy $(T_w)_{\mu\nu}$ localized at the wall. This is the most general form of action we can add to the wall and should easily lead to a solution. Because we don't want the boundary current or stress tensor to enter into the equations of motion at zeroth order, we have assumed that these terms enter the action at $O(\epsilon^2)$. Solving for these functions is straightforward, but tedious. We content ourselves with the fact that it can be done and move on to more interesting physics.

\end{enumerate}

\section{Vortex physics}\label{sec:thermo}
In this section, we analyze several aspects of the physics in the presence of the vortex lattice. We compute the vortex contribution to the scaling of thermodynamic quantities, and we discuss lattice formation in backgrounds with a smooth horizon. Finally, we discuss the correction to entanglement entropy due to the vortex lattice.  

The dimension of $\rho_1(r)$ given in equation (\ref{eq:dim}) determines much of the physics due to the vortex lattice at leading order in $\epsilon$. First off, we would like to know in what parameter range our linearized expansion is valid. As described in \S~\ref{sec:second_order}, the corrected metric functions and fields which appear at zeroth order have the form,
\be
f_i\sim f_i^0(1+c_i\epsilon^2 r^{2\Delta}+\ldots)
\ee
with $\Delta$ given by equation (\ref{eq:dim}) and $c_i$ some $O(1)$ constant. The solution will be valid for all ranges of $r$ such that the second term is subleading, i.e.
\be
r \ll \epsilon^{-{1\over \Delta}},
\ee
or, in terms of the cutoff scale, the expansion parameter must be
\be
\epsilon \ll r_0^{-\Delta}.
\ee
This means that, in the case of a hard wall cutoff at $r=r_0$, for a fixed choice of $r_0$ and $\Delta$, we can always find a sufficiently small value of $\epsilon$ such that the linearized approximation is valid for the entire range of the bulk. For $\theta=0$ this bound is
\be
\epsilon \ll r_0^{-\left (\frac{1}{2} \sqrt{{2m^2\over (z+2)(z-1)}+(z+2)^2}+1+{z\over 2}\right ) };
\ee
otherwise, it is
\be
\epsilon \ll r_0^{\theta-z-2}.
\ee
 Thinking of the parameter $\epsilon$ as the distance away from the critical point, we see that if we increase $\theta$, the linearized expansion is valid for a larger range of $\epsilon$ away from the critical field, whereas the converse is the case for increasing $z$. Equivalently, if we consider increasing $\theta$ with all other parameter values fixed, the linearized approximation will be valid over a larger range of $r$ in the bulk, whereas increasing $z$ decreases the value of the bulk radius at which the linearized approximation breaks down. This makes sense in light of the fact that $\theta$ has the effect of decreasing the dimensions of operators \cite{theta}, meaning the strength of the lattice will grow less quickly in the IR with a larger $\theta$. On the other hand, if one breaks Lorentz invariance with $z>1$, the scaling dimension of operators increases, and the relevance of the lattice grows more quickly into the IR. Thus, we expect the effects of the vortex lattice to be stronger in non-Lorentz invariant theories.

Assuming we are in this range for the entire bulk, we can calculate the parametric scaling of  the lattice contribution to thermodynamic quantities with the cutoff scale. The free energy in the field theory is given by evaluating the bulk on-shell action as
\be
F\sim TS_{on-shell}
\ee
where $T$ is an energy scale. In the hyperscaling geometries with a hard wall which we have considered above, the bulk is at zero temperature, as there is not a horizon, nor are there gapless excitations. However, the location of the hard wall still provides an energy scale, similar to the confinement scale $\Lambda_{QCD}$ in QCD because the field theory flows to a gapped phase in the IR. The location of the wall sets this energy scale as $\Lambda \sim {1\over r_0}$, as discussed in \S~4 of \cite{vlat}. Therefore, in analogy with systems with a horizon located at $r=r_0$, the free energy can be computed as a function of ``temperature" by evaluating the on-shell action as a function of $r_0$, and substituting $r_0\sim T^{-{1\over z}}$ to get temperature dependence\footnote{Let us run with the hard wall/black hole analogy for a bit longer. As described in \cite{theta}, for finite temperature Lifshitz spacetimes, the temperature of a black hole scales with the horizon location as $T\sim r_h^{-z}$.}. In \S~\ref{sec:temp}, we will discuss to what extent our results here for $O(\epsilon^2)$ vortex lattice physics extend to bona fide black hole geometries with a horizon, which have very different zeroth order IR physics.

Let's forestall any further concerns with this motivation, and determine the scaling of the free energy. Calculating the on-shell action to $O(\epsilon^2)$ is straightforward but tedious; we can satisfy ourselves by picking out the scaling of the action with $r_0$---this will give us the naive temperature dependence of the free energy.  We find that the dependence is
\be
F\sim T^{1+{2-\theta\over z}}\left(1+\epsilon^2 T^{-{2\Delta\over z}}+\ldots\right),
\ee
with $\Delta$ the correction due to the vortex lattice, given by equation (\ref{eq:dim}).

\subsection{Impurity thermodynamics}

We have seen the impurity contribution to the free energy in the dual field theory above. What else can we say about thermodynamics of the vortex defects? From our knowledge of the free energy, we can also determine the effect of the vortex lattice on the thermal entropy density, 
\be
S\sim {\partial F\over\partial T}\sim T^{{2-\theta\over z}}\left(1+\epsilon^2 T^{-{2\Delta\over z}}+\ldots\right)
\ee
 and specific heat density,
 \be
 C\sim -T{\partial^2F\over\partial T^2}\sim -T^{{2-\theta\over z}}\left(1+\epsilon^2 T^{-{2\Delta\over z}}+\ldots\right).
 \ee
 At $O(\epsilon^2)$ the vortices are non-interacting, so we can interpret the above quantities as the temperature dependence of the entropy and specific heat densities associated to one vortex.

\subsection{Backgrounds with a horizon}\label{sec:temp}
Now we will consider to what extent the physics of vortex lattice formation in systems with a hard wall cutoff carries over to lattice formation in black brane backgrounds.
Let's consider the backgrounds with a horizon and general critical exponents. The zeroth order form of the metric is now,
\be
ds^2=L^2r^{\theta-2}\left (-r^{2-2z}f(r)dt^2+\frac{dr^2}{f(r)}+dx^2+dy^2\right ),
\ee
where the emblackening factor is $f(r)=1$ at zero temperature and $f(r)=1-\left (\frac{r}{r_h}\right )^{2+z-\theta}$ at finite temperature, with temperature set by the horizon location, $r_h$, as $T\sim r_h^{-z}$. These metrics can be supported by Einstein gravity with a scalar and gauge field \cite{theta}. Once again, the NEC constrains the values of the critical exponents as in \S~\ref{sec:EMD}.

The scaling dimension of the vortex lattice operator is given by the radial equation for $\psi$ as before. At first order in $\epsilon$ this equation becomes,
\be\label{eq:thermal_psi}
f(r)\rho''(r)+\frac{1}{r}\left(\theta-z-1-\left ({r\over r_h}\right )^{1+z-\theta}\right )\rho'(r)-(m^2L^2r^{\theta-2} -\lambda_n)\rho(r)=0.
\ee
From this equation it is straightforward to see that near the boundary, $r\sim 0$, $\psi$ behaves as $r^\Delta$, with $\Delta$ given by equation (\ref{eq:dim}). Therefore, all of the near-boundary scaling results from above carry over to the case with a horizon. 

We can see this explicitly in the following example. For $\theta=0, z=1$, (i.e. Schwartzschild-$AdS_4$,) equation (\ref{eq:thermal_psi}) can be solved exactly. The solution is 
\be
\rho(r)= \left ({r\over r_h}\right )^{{\gamma\over 2}} ~_2F_1\left({\gamma-2\sqrt{Q}r_h\over 4},{\gamma+2\sqrt{Q}r_h\over 4},{\gamma-1\over 2},{r^2\over r_h^2}\right),
\ee
where $\gamma=3+\sqrt{9+4m^2L^2}$ and $_2F_1$ is the hypergeometric function. Near the boundary this scales as $r^{\gamma\over 2}$, as expected, just as in the case with a hard wall. Near the horizon, the scaling is 
\be
\psi\sim (r-r_h)^{-{1\over 2}},
\ee
indicating the linearized approximation is breaking down as $r\to r_h$. Therefore, as long as we stay in the regime where the linearized approximation is valid, the scaling dimension of the lattice is given by the near boundary expansion, which is the same in both the case of a hard wall and a horizon. This expression also tells us that the lower the temperature, (the larger the value of $r_h$,) the larger the range of $r$ in the bulk for which the linearized approximation is valid. However, unlike in the case of the hard wall, the linearized approximation can never be valid all the way to $r=r_h$, at which point one must find a nonperturbative solution.

Now, let's consider the set of parameters  $\theta=1$ and $z=2$. After rescaling $r\to r/r_h$, we get
\be
(1-r^3)\rho''(r)-\frac{1}{r}(2+r^3)\rho'(r)-\left (\frac{m^2r_h}{r}-Q r_h^2\right)\rho(r)=0,
\ee
We cannot solve this analytically, but we can expand it in a power series near the boundary and near the horizon. This will allow us to determine the critical value of $r_h$ at which there first exists a normalizable solution for $\rho(r)$. This is the critical temperature at which the droplet condensate begins to form.

In order to find the critical temperature, we will employ the following near-boundary expansion
\be\label{eq:boundary_exp}
\rho_b(r)=\sum_{n=0}^\infty \alpha_nr^n,
\ee
and the near-horizon expansion
\be\label{eq:horizon_exp}
\rho_h(r)=\sum_{m=0}^\infty \beta_m(1-r)^m.
\ee
First let's consider the boundary. We require that there is no non-normalizable mode of $\rho(r)$ at the boundary; this means we require $\rho(r)\to 0$ as $r\to 0$. With this requirement, the first non-zero constant is $\alpha_3$, and all of the higher constants in the expansion (\ref{eq:boundary_exp}) are proportional to $\alpha_3$. We require regularity at the horizon, which means as $r\to 1$, $\rho_h(r)\to \beta_0$ is a constant. We are free to set $\beta_0$ to whichever value we choose. All of the other $\beta_m$ in equation (\ref{eq:horizon_exp}) are proportional to this $\beta_0$. If we choose values for $\beta_0, m^2$, and $Q$, we have two free parameters, $\alpha_3$ and $r_h$, which we determine by requiring the near-boundary and near-horizon expansions and their derivatives to match at $r=1/2$, i.e. $\rho_b(r=1/2)=\rho_h(r=1/2)$ and $\rho_b'(r=1/2)=\rho_h'(r=1/2)$. 

\begin{figure}[htb]
\centering
\includegraphics{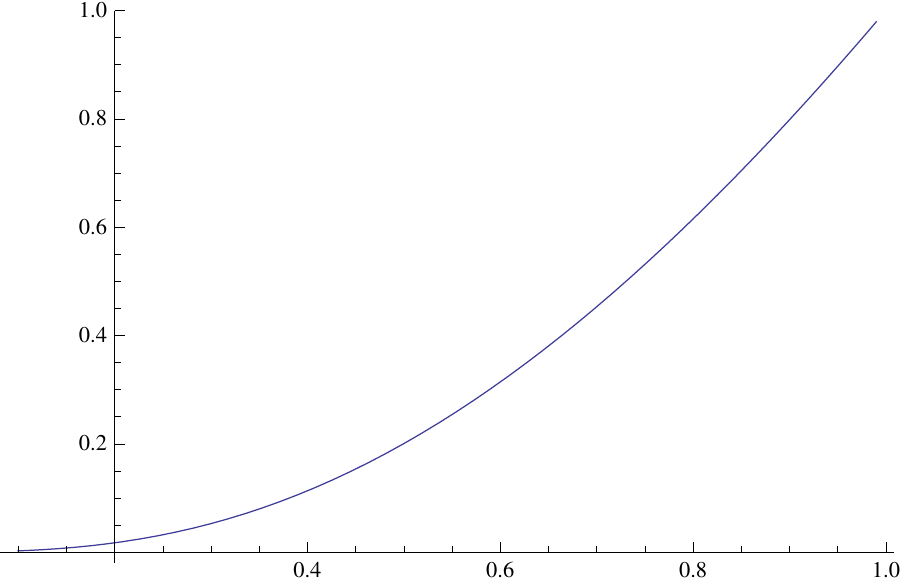}\caption{Solution for $\rho(r)$ at the critical temperature, $r_h\approx 1.70$. $r=0$ is the boundary and $r=1$ is the horizon.}\label{fig:rho_r}
\end{figure}

\begin{figure}[htb]
\centering
\includegraphics[width=3in]{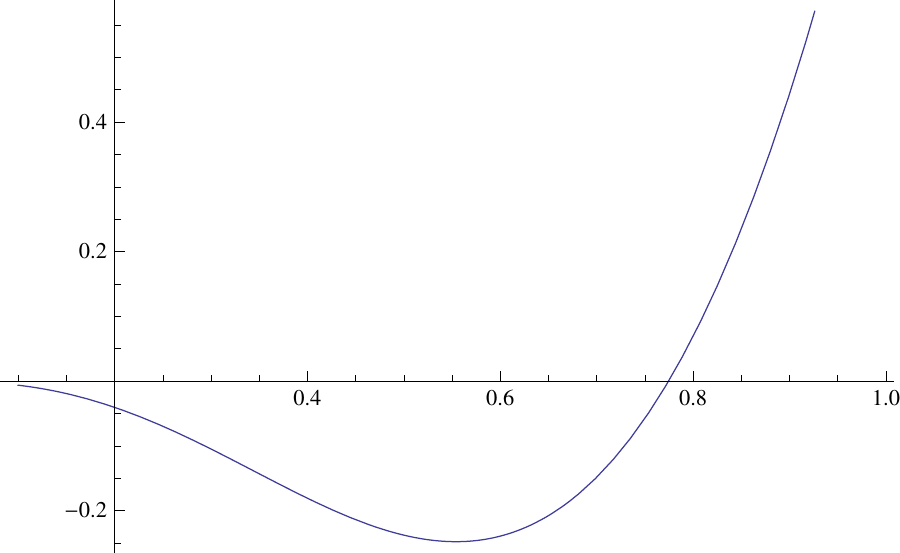}
\includegraphics[width=3in]{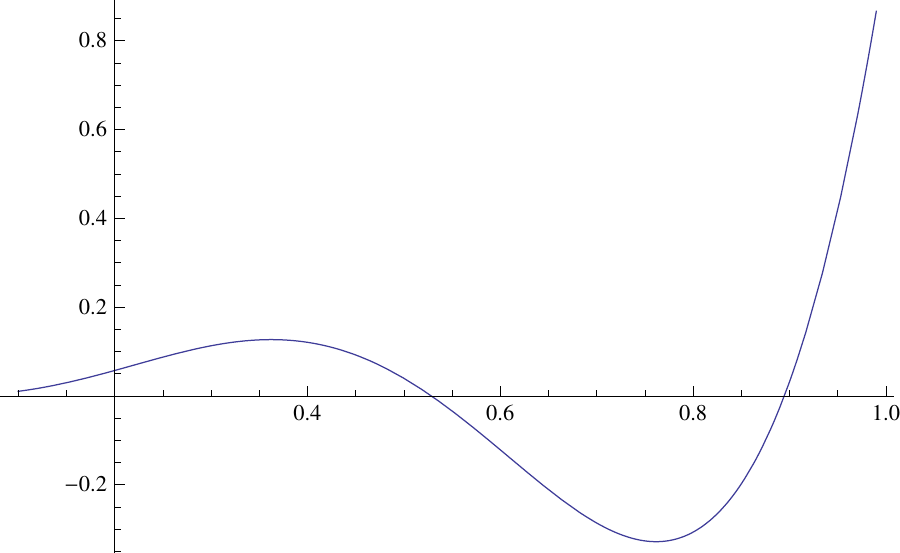}\caption{Left: First excited state solution for $\rho(r)$, with $(\alpha_3,r_h)\approx (-7.71,3.54)$. Right: Second excited state solution with $(\alpha_3,r_h)\approx (15.46,5.52)$}\label{fig:other_r_solns}
\end{figure}

We will determine the critical value of $r_h$ for the case $\beta_0=1$, $m^2=-2$, and $Q=1$. Expanding $\rho_b$ and $\rho_h$ to eleven terms each and imposing the matching conditions at $r=1/2$ gives us a set of pairs $(\alpha_3,r_h)$ for which the equations are satisfied. The critical temperature is the lowest value of $r_h$, which corresponds to the solution with no nodes. For our parameter choices, we get that the lowest order solution is given by $(\alpha_3,r_h)\approx (2.57,1.70)$. 
This solution is shown in Figure (\ref{fig:rho_r}). The next two higher order solutions are plotted in Figure (\ref{fig:other_r_solns}).

Once we have a horizon in play, many of the somewhat ad hoc results we found before become quite salient. We can interpret the $O(\epsilon^2)$ on-shell action near the boundary as the leading order lattice contribution to thermodynamic entropy.

We can also imagine that if one could compute the full nonperturbative backreacted lattice solution in these geometries, one would find a ``crystalline" horizon, i.e. a black brane with translation invariance broken to a discrete group.
\subsection{Entanglement Entropy}

We would like to compute the leading order contribution of the lattice to the scaling of entanglement entropy in the dual field theory with entanglement region size. To extract the scaling dimension for entanglement regions large compared to the lattice size, it is sufficient to use the metric obatined by coarse graining over the periodicity in $x$ and $y$. We will refer to this limit as the ``smeared limit;" it is valid for entanglement regions of size $\ell$ where $\ell$ is much larger than the lattice spacing. As our lattice has dimension $v_1\times {2\pi\over v_1}$, where $v_1=\sqrt{2\pi}$ for the square lattice, we require $\ell\gg O(1)$. 

We will do this for the case of lattice formation in the metric of equation (\ref{eq:metric}). Near the boundary, the metric coarse-grained over the lattice correction has the form
\be
ds^2=L^2r^{\theta}\left (-{1\over r^{2z}}(1-\epsilon^2 a r^{2\Delta})dt^2+{1\over r^2}(1+\epsilon^2 a_1 r^{2\Delta})dr^2+{1\over r^2}(1+\epsilon^2 br^{2\Delta})(dx^2+dy^2)\right ),
\ee
where $a, a_1$ and $b$ are constants, the zero-mode coefficients of $a(x,y), a_1(x,y)$ and $b(x,y)$, respectively, and $\Delta$ is the near-boundary scaling dimension of the vortex lattice correction. As the higher modes are exponentially suppressed, these are the only terms that matter. The action for a strip  in the $x-y$ plane with width $-{\ell\over 2}\leq x\leq{\ell\over 2}$, and length $-{L_y\over 2}\leq y\leq{L_y\over 2}$, $\ell\ll L_y$ is
\be
S=2L_y\int_{r_c}^\delta dr \sqrt{g_{rr}g_{xx}}\sqrt{1+{g_{xx}\over g_{rr}}x'^2},
\ee
where $r_c$ is the extremal value of the minimal surface in the bulk and $\delta\ll 1$ is a UV cutoff. 

We can approximate the $\ell$ scaling of the entanglement entropy using a method of trial surfaces. As discussed in \cite{theta}, such an approximation is sufficient to capture the correct scaling properties of the entanglement entropy, though the surface is not the strict minimum area surface. Instead of computing the minimal surface for the strip geometry, we will approximate the strip as a rectangular surface with two sides of dimension $-{L_y\over 2}\leq y\leq {L_y\over 2}, \delta \leq r\leq r_t$ localized at $x=\pm{\ell\over 2}$ and base of dimension $-{L_y\over 2}\leq y \leq {L_y\over 2},  -{\ell\over 2}\leq x\leq {\ell\over 2}$ localized at $r=r_t$. Then we will minimize the area of this surface with respect to the radius $r_t$ to which it extends in the bulk,
\be
\mathcal A\sim L^2L_y\left (2\int_{r_t}^\delta dr \sqrt{g_{rr}g_{xx}}+\ell g_{xx}|_{r=r_t}\right ).
\ee
The first term in the above expression comes from integrating the sides of the rectangle in the  $r-y$ plane, while the second term comes from integrating the bottom of the rectangle, which is an $L_y\times\ell$ surface localized at $r=r_t$. We want to minimize the area of this surface with respect to $r_t$. 

To do this, we differentiate $\mathcal A$ with respect to $r_t$ and set this equal to zero. This gives us the relation $r_t\sim \ell+\epsilon^2\ell^{2\Delta}$; plugging this back into the equation for the area, we get that the leading order scaling with $\ell$ is,
\be
\mathcal A \sim L^2L_y \ell^{\theta-1}(1+c\epsilon^2\ell^{2\Delta}+\ldots )
\ee
where $c$ is an $O(1)$ constant and $\Delta$ the dimension found in equation (\ref{eq:dim}). By checking the second derivative, we also see that the correction to the entanglement entropy is positive.

As expected, at zeroth order we find entanglement entropy which scales as a $d=2$ theory with hyperscaling violation exponent $\theta$ \cite{HSS,theta}, and at $O(\epsilon^2)$, it receives a $\Delta$-dependent correction due to the vortex lattice. As we discuss in the next section, we can view the correction due to the coarse-grained vortex lattice as producing hyperscaling violation. We note that the method of trial surfaces is known to fail for geometries which have an $AdS_2$ region \cite{swingle}, however numerical results show that at leading order the entanglement scales as an area law $\sim L_y\ell$, with a subleading correction due to the vortex lattice that is power law in $\ell$. 

\subsection{Dynamical generation of hyperscaling violation}
In this section we will provide evidence  that vortex impurities contribute to the physics as if each one is a (0+1)-dimensional system with an effective $\theta$ between 0 and 1. Here we demonstrate this for the scenario discussed in \cite{vlat}, where the background geometry at zeroth order is $AdS_2\times R^2$, but it is straightforward to see that similar results carry over to the cases discussed in \S~\ref{sec:gen} with general $z,\theta$. 

We will see evidence of this in two ways: 1) the impurity contribution to the free energy,  and 2) the (smeared) vortex lattice correction to the scalar curvature. In each of these cases, the vortex contribution has the scaling of a theory with $\theta = 2\alpha$, which, because ${1\over 2}<\alpha<1$, is a theory with hyperscaling violation exponent between $d-1 (=1)$ and $d(=2)$.

\begin{itemize}
\item Free energy:

We have already seen that the free energy to $O(\epsilon^2)$ scales as
\be
F\sim T(1+\epsilon^2T^{-2\alpha}+\ldots)
\ee
after spontaneous lattice generation in $AdS_2\times R^2$. Let's compare the scaling of the correction at $O(\epsilon^2)$ to the scaling of free energy in a geometry with $z=1$ and hyperscaling violation exponent\cite{theta},
\be
F\sim T^{1+\frac{(d-\theta)}{z}}\sim T^{1-\theta}.
\ee
Using the fact that each impurity has $d=0$ in our case, we have \cite{vlat} 
\be
T^{1-2\alpha}\sim T^{1-\theta}\implies \theta=2\alpha.
\ee

\item Scalar curvature:

It is straightforward to see the effect of the vortex lattice on the scalar curvature at distances much larger than the lattice spacing.  A simple computation of the Ricci scalar for the smeared limit discussed in the previous section yields,
\be
R\sim -{2\over L^2}(1+c\epsilon^2r^{2\alpha}+\ldots),
\ee
where we note that $-{2\over L^2}$ is the scalar curvature for $AdS_2\times R^2$. 
Comparing this to the scalar curvature for theories with hyperscaling violation, $R\sim r^\theta$, we see that at $O(\epsilon^2)$,
\be
R\sim r^\theta \implies \theta = 2\alpha,
\ee
which is the same result as found above. 
\end{itemize}

 At $O(\epsilon^2)$, the lattice of (0+1)-dimensional defects forms a (2+1)-dimensional theory with effective hyperscaling violation exponent $\theta=2\alpha$. Increasing the value of the scalar mass or the background magnetic field decreases the value of $\alpha$, which causes the strength of the lattice to grow less quickly. As argued in \cite{theta}, $d-\theta$ can be viewed as the effective (spatial) dimensionality of our theory; this implies that our effective dimensionality is $2-2\alpha$, a value between 1 and 2. Thus the presence of the vortex lattice results in a fractional dimension which is less than the naive dimension in the UV; such behavior was found by a holographic entanglement entropy analysis of hyperscaling violating metrics in section 4 of \cite{theta}, and explicitly constructed in a lattice system in \cite{SW}. 

In \cite{vlat}, we conjectured that as one flows to the IR of the theory, the effect of the vortex lattice grows so large that the geometry splits into a collection of independent $AdS_2$s in a manner similar to what is discussed in \cite{fragment}. Though we have found the appearance of hyperscaling violation at subleading order in the linearized approximation, we may conjecture that the full nonperturbative vortex lattice solution would exhibit a metric which has an intermediate IR region with effective hyperscaling violation exponent $\theta=2\alpha$, before the theory fully splits into separate $AdS_2$ regions. We can think of the appearance of hyperscaling violation in this intermediate IR region as indicating that the effective dimensionality of the theory is decreasing, interpolating between the (2+1)-dimensional UV fixed point and the effectively (0+1)-dimensional IR fixed point. It would be very interesting if one had sufficient muscle to see this nonperturbatively.
\section{Discussion}\label{sec:discussion}
In this work we have extended the results of \cite{vlat} to systems with general Lifshitz and hyperscaling violation exponents which transition to a vortex lattice phase.  We have also found such phases in systems with smooth black brane horizons. In both cases we find that in the linearized approximation the vortex lattice phase can persist for a substantial range of $r$ in the bulk. As long as $r \ll \epsilon^{-{1\over\Delta}}$, where $\epsilon$ is the distance from the critical point and $\Delta$ is a positive power, the linearized approximation remains valid.

We also consider the entanglement entropy, free energy, and scalar curvature of the smeared vortex lattice solution and find that the $O(\epsilon^2)$ corrections generically scale as if they had a positive hyperscaling violation exponent. Indeed in the $AdS_2\times R^2$ case we find a hyperscaling violation exponent with $d-1<\theta <d$ which is tunable with the mass of the scalar introduced, allowing the exploration of a natural realization of such hyperscaling exponent violating theories discussed in \cite{theta}. We can consider the appearance of a hyperscaling violating phase in the intermediate IR region as signaling a decrease in the effective spacetime dimensionality of the theory, interpolating between a (2+1)-dimensional conformal fixed point in the UV and an effectively (0+1) dimensional conformal fixed point in the IR, where the individual lattice sites have completely decoupled and the spacetime has fragmented into individual $AdS_2\times R^2$ regions \cite{fragment}.

There are a number of interesting directions one could pursue in the systems we have studied here. Firstly, we should point out that we have only established that there is a critical scale dimensionless in $B$ and the location of the hard wall at which the square lattice is preferable to the uncondensed phase;  it would be interesting to check which lattice shape is the most preferred. 
 in order to do this we would need to carry out the linearized backreaction on $\psi, A_\mu$, and $g_{\mu\nu}$ to $O(\epsilon^4)$. This would also allow us to see find the minimum of the free energy for different lattice shapes as a function of $z$ and $\theta$. We could then compare the critical $B/T^2$ for different values of critical exponents, and determine if the lattice generically condenses before or after one hits the $AdS_2\times R^2$ region in the IR. It could be that the preferred lattice shape is something that changes as some subset of $z$, $\theta$, the temperature, the magnetic field strength, and either the location of the hard wall or the presence of a horizon is varied, depending on the construction. Such an analysis would generically have to be numerical.

We now list a number of other, more ambitious directions.
\begin{itemize}
\item 
It would be interesting to numerically study the case of the smooth black hole solutions fully, past the point where perturbation theory in $\epsilon$ breaks down. Such an analysis would require numerical machinery that is past the scope of this paper.

\item One would like to check the current-current correlation function in the presence of the lattice phase to see if it is insulating, as one would expect from the conjectured solid phases which emerge in the IR of doped CFTs \cite{Sachdevnew}.

\item
It would also be interesting to check the degree of metastability of the lattice by checking the phonon modes thereof. This is potentially also interesting in the context of the lattice shape, as it could be that the potential minima associated with specific lattice shapes are shallow and generically unstable as local minima.

\item 
We have so far only studied instabilities of a complex scalar coupled to the magnetic field. It would be interesting to study fermionic instabilities in the backgrounds we have discussed.

\item
It could also be possible to use a theory similar to the one considered in this paper to model the transition from Type II superconductivity into a vortex lattice phase. It is already clear from both this and \cite{vlat} that the size of the superconducting droplets in the vortex lattice model is something that is controlled by both temperature and magnetic field; it therefore may be possible to marry the existing results on holographic superconductors with our vortex lattice result to holographically describe the transition between a normal superconductor and a vortex lattice phase.

\item
Finally, it is possible that our vortex lattice could melt into a vortex glass or even a liquid phase at higher temperatures. Any study on such a possibility would probably have to be a numerical one.

\end{itemize}

\bigskip
\centerline{\bf{Acknowledgements}}
We would like to thank X. Dong, S. Hartnoll, S. Kachru, E. Shaghoulian, and G. Torroba for discussions. We also thank N. Paquette for comments on a draft. SH is supported in part by the John Templeton Foundation.
\appendix

 \section{Linearized equations of motion at $\epsilon^2$ order}
 We give the equations of motion for the correction to the metric functions, dilaton, and gauge field at $O(\epsilon^2)$ in the linearized approximation, as discussed in \S~\ref{sec:second_order}. We employ the expansions of equation (\ref{eq:exps}) for the functions, whose values are sourced by terms which scale as $\sim|\psi_1|^2$. The following are the ODEs for the $r$-dependent Fourier coefficients in equation (\ref{eq:dim}) in the case of general $z$ and $\theta$ with $z\neq 1$. We do this for $Q=1$ and $v_1=\sqrt{2\pi}$, which is a square lattice. We have suppressed the parameters; i.e. $\tilde f_i(r,k,l)\equiv f_i,$ $\forall f_i\in  \{a,a_1,b,a_2^x,a_2^y,\phi_2 \}$.
 
 We find it convenient to define the constants,
 \be
 \gamma_1\equiv \sqrt{\theta^2-2z(\theta-2)-4},~\gamma_2\equiv (z-1)(2+z-\theta),
 \ee
 which are nonzero for generic values of $z,\theta$, with $z\neq 1$.
 
 Now we will give the equations of motion, and then discuss how they can be solved for general values of $(k,l)$ in the Fourier expansion. Primes denote derivatives with respect to $r$, and $\rho_1(r)$ is the radial solution for the complex scalar as found in \S~\ref{sec:psi_sol}.

First we give the gauge field equations of motion for $A_x$ and $A_y$ respectively,
 \begin{multline}
 \sqrt{2 \pi } l  (a_1-a-2b)   +\frac{2  \sqrt{2 \pi } (\theta -4) l 
   \phi_2 }{\gamma_1}
   +\frac{\sqrt{\pi} l r^2\rho_1^2}{\gamma_2}\\
   =i\left (4 \pi   l  (ka_2^y-la_2^x)+2{\theta -3 - z\over r}
   (a_2^x)'+2  (a_2^x)''
    \right ),
 \end{multline}
 and
 \begin{multline}
 \sqrt{2 \pi } k  (a-a_1+2b)  -\frac{2  \sqrt{2 \pi } (\theta -4)
   k  \phi_2 }{\gamma_1}
   -\frac{ \sqrt{\pi} kr^2 \rho_1^2}{\gamma_2}\\
   =i\left ( 4 \pi  k   (la_2^x-ka_2^y)
 +2{\theta -3  - z
   \over r}(a_2^y)' +2  (a_2^y)''  \right ).
 \end{multline}
 From these we see that the zero modes ($k=l=0$) are vanishing. Additionally, we have the constraint
 \be
k(a_2^x)'+l(a_2^y)'=0, 
 \ee
 which comes from the equation of motion for $A_r$. Together these provide one independent equation.
 
Next we give the equation of motion for the dilaton,
 \begin{multline}
-\frac{2 \gamma_1 a_1 }{z-1}+\frac{r \gamma_1(a'+a_1'-2b') }{\gamma_2}
   +{4 (\theta -4) \over\gamma_1}  i \sqrt{2 \pi }(la_2^x -ka_2^y) \\
   +\frac{4 \pi  (k^2+l^2) r^2  \phi_2 }{\gamma_2}+\frac{2(\theta-2)( \theta ^2-8(z-1)) \phi_2 }{\gamma_1^2(z-1)}
   -\frac{2r^2 \phi_2'' }{\gamma_2}+\frac{2 r (z+1-\theta)  \phi_2' }{\gamma_2}=0.
 \end{multline}
 Note that all the dilaton does not directly couple to $\psi_1$, it is still affected by the lattice through its coupling to the gauge field and metric, which introduce terms in its equation of motion at $O(\epsilon^2)$.
 
Finally, we list the nontrivial Einstein equations.
   Equation for $G_{tt}$:
  \begin{multline}
  2\pi r^2(k^2+l^2)(a_1+b)-\gamma_2{\theta-2\over z-1}a_1+2i\gamma_2\sqrt{2\pi}(ka_2^y-la_2^x)+2\gamma_2b+{\gamma_1\gamma_2\over z-1}\phi_2\\+r((\theta-2)(a_1'-2b')-\gamma_1\phi_2')-2r^2b''={1\over \sqrt{2}}\left (r^2\rho_1^2(1-\pi(k^2+l^2))+r^2\rho_1'^2+{m^2r^{\theta}\over 2\gamma_2}\rho_1^2\right )
  \end{multline}
   
   Equation for $G_{rr}$:
  \begin{multline}
  2\pi r^2(k^2+l^2)(a-b)+\gamma_2a_1+2i\gamma_2\sqrt{2\pi}(la_2^x-ka_2^y)-2\gamma_2b-{\gamma_1\gamma_2\over z-1}\phi_2\\+(2-\theta)ra'+2(\theta-z-1)rb'-\gamma_1\phi_2'={1\over\sqrt{2}}\left (-r^2\rho_1^2(1-\pi(k^2+l^2))+r^2\rho_1'^2-{m^2r^{\theta}\over 2\gamma_2}\rho_1^2\right )
  \end{multline}
    
     Equation for $G_{rx}$:
 \begin{equation}
 \sqrt{2\pi}\gamma_2i(a_2^y)'+kr\pi(r(b'-a')+\gamma_1\phi_2-(1-z)a-(\theta-z-1)a_1)=-{\pi k\over\sqrt 2}r^2\rho_1\rho_1'
 \end{equation}
    
 Equation for $G_{ry}$:
 \begin{equation}
 \sqrt{2\pi}\gamma_2i(a_2^x)'+lr\pi(r(a'-b')-\gamma_1\phi_2+(1-z)a+(\theta-z-1)a_1)={\pi l\over \sqrt 2}r^2\rho_1\rho_1'
 \end{equation}
    
Equation for $G_{xx}$:
\begin{multline}
2\pi l^2 r^2 (a-a_1) + 2\sqrt{2\pi} i\gamma_2(ka_2^y-la_2^x)-(2z^2+4z-3\theta z+\theta(\theta-2))a_1+2\gamma_2b-{\gamma_1\gamma_2\over z-1}\phi_2\\+r((2z-\theta)a'+(1+z-\theta)(a_1'-b')+\gamma_1\phi_2')+r^2(b''-a'')=-{1\over \sqrt 2}\left(r^2\rho_1'^2 +{m^2r^\theta\over 2\gamma_2}\rho_1^2\right )
\end{multline}
    
Equation for $G_{yy}$:
\begin{multline}
2\pi k^2 r^2 (a-a_1) + 2\sqrt{2\pi} i\gamma_2(ka_2^y-la_2^x)-(2z^2+4z-3\theta z+\theta(\theta-2))a_1+2\gamma_2b-{\gamma_1\gamma_2\over z-1}\phi_2\\+r((2z-\theta)a'+(1+z-\theta)(a_1'-b')+\gamma_1\phi_2')+r^2(b''-a'')=-{1\over \sqrt 2}\left(r^2\rho_1'^2 +{m^2r^\theta\over 2\gamma_2}\rho_1^2\right )
\end{multline}

Constraint from $G_{xy}$:
\be
kl(a-a_1)=0.
\ee
 This means that the functions $a(r,k,l)$ and $a_1(r,k,l)$ are the same in all cases where $k,l\neq 0$. (This is also implied by considering $G_{xx}-G_{yy}$.) Therefore, there is only one independent equation and a constraint in the set of equations for $G_{xx},G_{yy},$ and $G_{xy}$.
 
 All in all, we have six independent differential equations,\footnote{The set of equations for $\{A_x,\phi,G_{tt},G_{rr},G_{rx},G_{xx}\}$ is one linearly independent set.} which we can in principle decouple in order to get a second order differential equation for each function. The UV boundary condition is that all functions are normalizable. Using a scaling argument we  see that they all take the form $\sim r^{2\Delta}$ near the boundary, where $\Delta$ is given in equation (\ref{eq:dim}). The IR boundary condition is set via the source of stress energy located at the hard wall, as described above.

\end{document}